\documentclass[hyper]{JHEP} 

\usepackage{epsfig}

\newcommand\fverb{\setbox\pippobox=\hbox\bgroup\verb}
\newcommand\fverbdo{\egroup\medskip\noindent%
			\fbox{\unhbox\pippobox}\ }
\newcommand\fverbit{\egroup\item[\fbox{\unhbox\pippobox}]}
\newbox\pippobox

\title{Remark about Non-BPS D-Brane in  Type IIA Theory}

\author{by J. Kluso\v{n}\\
	 Department of Theoretical Physics and Astrophysics\\
                   Faculty of Science, Masaryk University\\
Kotl\'{a}\v{r}sk\'{a} 2, 611 37, Brno\\
Czech Republic\\
	E-mail: \email{klu@physics.muni.cz}}

\preprint{\hepth{9909194}}	

\abstract{In this paper we would like to show simple mechanisms how from
the action for non-BPS Dp-brane  we can  obtain action
describing BPS D(p-1)-brane in IIA theory.}


\begin{document}

\section{Introduction}

In the last year many new results about D-branes in string theories have emerged.
In the remarkable series of paper by Sen \cite{Sen1,Sen2,Sen3,Sen4,Sen5}, the problem of
 nonsupersymmetric configuration in string theories has been studied.   It is
clear that non-BPS D-branes in the IIA, IIB theories are as important as supersymmetric
ones. On the other hand it is known that non BPS D-branes are not stable, so that they can decay into supersymmetric
string vacuum. Instability of this system is a consequence of tachyon field
 that lives on world-volume of
non-BPS D-brane. But presence of non-BPS brane in the string theory is important from one 
reason. We can construct  kink solution of tachyonic field on the world-volume of non-BPS D-brane, which 
forms D-brane of dimension smaller than original non-BPS D-brane. It can be shown
on base of topological arguments that this solution is stable (for a review of this subject, see \cite{Schwarz,SenA}).
Witten generalised this construction and showed that all branes in IIB theory can be
constructed as topological defects in space-time filling world-volume of D9 branes and D9 antibranes
\cite{witen}. Ho\v{r}ava extended this construction to the case of IIA theory and  showed
that all D-branes in IIA theory can emerge as topological solutions in space-time filling
non-BPS D9 branes . Ho\v{r}ava also proposed an intriguing conjecture about Matrix theory
and  construction of D0 branes in K theory \cite{Horava,Horava1}. For review of the subject D-branes
and K theory, see \cite{Olsen2}, where many references can be found.

In recent paper Sen \cite{Sen} proposed an supersymmetric invariant action for non-BPS D-branes.
Because non-BPS branes break all supersymmetries, it seams to be strange to construct
supersymmetric action describing this brane. However, although there is no manifest 
supersymmetry of world-volume theory, we still expect world-volume theory to be supersymmetric,
with the supersymmetry realised as a spontaneously broken symmetry.
From these arguments Sen showed that the action  has to contain the full number of fermionic zero modes (32),
because they are fermionic Goldstone modes of completely broken supersymmetry, while BPS D-brane
contains   16 zero modes, because breaks one half of supersymmetry. Sen showed that the DBI action for
non-BPS D-brane (without presence of tachyon) is the same as the
supersymmetric action describing BPS D-brane.
This action is manifestly symmetric under all space-time supersymmetries. Sen argued
that ordinary action for BPS D-brane contain DBI term and WZ term, which are invariant
under supersymmetry but only when they both are present in the action for D-brane, the action is  invariant under local symmetry on brane $ \kappa $
symmetry that is needed for gauging away one half of fermionic degrees of freedom, so that on BPS D-brane only
16 physical fermionic fields live, as should be for object breaking 16 bulk supersymmetries. Sen showed that  the DBI term for non-BPS D-brane is exactly the same 
as  DBI term in action of BPS D-brane (when we suppose, that other massive fields are integrated out, including tachyon)
that is invariant under sypersymmetric transformations, but has no a $ \kappa $ symmetry,
so that number of fermionic degrees of freedom is 32 which is a appropriate number of
fermionic Goldstone modes for object braking bulk supersymmetry completely.

Sen also showed how we could include the tachyonic field  into action. Because mass
of tachyon is of order of string scale, there is not any systematic way to construct
this effective action for tachyon, but on grounds of invariance under
supersymmetry and general covariance Sen proposed the form of this term expressing
interaction between tachyon and other light fields on world-volume of non-BPS D-brane.
 This term has a useful property
that for constant tachyon field is zero, so that the action for non-BPS vanishes identically.

In present paper we would like to extend the analysis in paper \cite{Sen}. We propose
the form of the term containing the tachyon and we show that condition of invariance under
supersymmetric transformation places strong constraints on the form of this term. Then we
will show that tachyon condensation in the form of kink solution leads to the DBI action
for BPS D-brane of codimension one with gauged local  $\kappa$ symmetry.

In conclusion, we will discuss  other  problems with non-BPS D-brane and relation of this construction
to the K-theory.

\section{Action for non-BPS D-brane in IIA theory}
We start this section with recapitulating basic facts about non-BPS D-branes in IIA theory, following \cite{Sen}.
Let $ \sigma_{\mu} , \ \mu=0,...p $ are world-volume coordinates on D-brane. Fields living
on this D-brane arise as the lightest states from spectrum of open string ending on this D-brane. These open strings have two CP sectors \cite{Sen2}: first, with unit $ 2\times 2 $ matrix, which correspond
to the states of open string with usual GSO projection $ (-1)^{F}\left| \psi \right>=\left| \psi \right > $, 
 where $ F$ is world-sheet fermion number and $ \left| \psi \right> $ is state from Hilbert space of
open string living on Dp-brane. The second CP sector has CP matrix  $ \sigma_1 $ and contains states
having opposite GSO projection $ (-1)^{F} \left| \psi \right>=-\left| \psi \right> $. The massless fields living
on Dp-brane are ten components of $ X^M(\sigma), M=0,...,9 $ ;  $ U(1) $ gauge field $ A(\sigma)_{\mu} $ and
 fermionic field $ \theta $ with $32$ real components transforming as Majorana spinor under transverse Lorenz
group $ SO(9,1) $. We can write $ \theta $ as sum of left handed Majorana-Weyl spinor and right handed
Majorana-Weyl spinor:
\begin{equation}
\theta=\theta_L+\theta_R ,\ \Gamma_{11}\theta_L=\theta_L, \ \Gamma_{11}\theta_R=-\theta_R
\end{equation}
All fields except $ \theta_R $ come from CP sector with identity matrix, while $\theta_R $ comes from 
sector with  $ \sigma_1 $ matrix. 
\footnote{Our conventions are following. $\Gamma^M$ are $32\times 32$ Dirac matrices
appropriate to 10d with relation $\{\Gamma^M,\Gamma^N\}=2\eta^{MN}$ with 
$\eta^{MN}=(-1,0,...,0)$. For this choice of gamma matrices the massive Dirac equation
is $(\Gamma^M\partial_M-M)\Psi=0$. We also introduce $\Gamma_{11}=
\Gamma_0...\Gamma_9 , (\Gamma_{11})^2=1$. We also work in units $4\pi^2\alpha'=1$.
Then tension of BPS Dp-brane is equal to $T_p=\frac{2\pi}{g}$ where $g$ is
a string coupling constant.}

As Sen \cite{Sen} argued, action for non BPS D-brane (without tachyon) should go to the action for BPS D-brane, when
we set $ \theta_R=0 $ (we have opposite convention that \cite{Sen}). From this reason, action
for non-BPS D-brane in \cite{Sen} has been constructed as supersymmetric  DBI action, which is manifestly
supersymmetric invariant, but has not $ \kappa $ symmetry, so we cannot gauge away one half of fermionic
degrees of freedom, so that this action describes non-BPS D-brane.

The next thing is to include the effect of tachyon. In order to get some relation between tachyon condensation and
supersymmetric D-branes, we would like to have an effective action for massless field and tachyon living on world-volume of  non-BPS D-brane. This effective action should appear after integrating out all massive modes of open string ending on Dp-brane. Because tachyon  mass is of order of string scale,
there is no systematic way to obtain effective action for this field, but we can still study some general properties of 
this action. Following ref.\cite{Sen}, the effective action for non-BPS Dp-brane with tachyonic field on its
world-volume should has a form:

\begin{equation}\label{action}
 S=-\int d^{p+1}\sigma\sqrt{-\det (\mathcal{G}_{\mu\nu}+\frac{1}{2\pi}\mathcal{F}
_{\mu\nu} )}F(T,\partial T, \theta_L,\theta_R,\mathcal{G},..) 
\end{equation}

\begin{equation}
\Pi^M_{\mu}=\partial_{\mu}X^M-\overline{\theta}
\Gamma^M\partial_{\mu}\theta \ , \ \mathcal{G}_{\mu\nu}=
\eta_{MN}\Pi^M_{\mu}\Pi^N_{\nu} 
\end{equation}
and
\begin{equation}
\mathcal{F}_{\mu\nu}=F_{\mu\nu}-[\overline{\theta}
\Gamma_{11}\Gamma_M\partial_{\mu}\theta 
(\partial_{\nu}X^M-\frac{1}{2}\overline{\theta}\Gamma^M
\partial_{\nu}\theta) -(\mu \leftrightarrow \nu ) ]
\end{equation}

The constant  
$ C=\sqrt{2}T_p=\frac{\sqrt{2}2\pi}{g} $ was included in the function $ F $  
in(\ref{action}).   

We must say few words about function $F$, which expresses
the presence of tachyon on world-volume of unstable non-BPS D-brane.
We know that this function must be invariant
under Poincare symmetry and supersymmetry. We also expect, that this
function must express interaction between light massless fields living
on world-volume of non-BPS D-brane and tachyon. We will also suppose in
construction of this function, that other massive fields were integrated out.
And finally, following \cite{Sen} we demand, that this function is zero
for tachyon equal to its vacuum expectation value $T_0$ and for $T=0$ is equal to 
$T_p$ tension of BPS D-brane, which corresponds to $(-1)^{F_L}$ projection, which
projects out tachyon field and also fermionic field from $\sigma_1$ sector, so that
resulting D-brane is BPS Dp-brane in Type IIB theory \cite{SenA}
\footnote{In fact, Sen  argued, that in case of constant $ T $,  $ F $ reduces to the potential for tachyon and as a 
consequence of general form of potential for tachyon, this term is zero for $ T=T_0 $. In this article, we slightly change behaviour
of this function, because we only demand that in point on world-volume, where the tachyon is equal to its vacuum value $ T_0 $, we
should recover supersymmetric vacuum, so that there are no fields living on non-BPS D-brane, so we have (\ref{c}). 
Sen also argued that for $T=0$ function $F$ should be equal to $C$, but we think that this function should 
rather be equal to $T_p$ from
reasons explained above.}
:
\begin{equation}\label{c}
F(T=T_0)=0 , \ F(T=0)=\frac{2\pi}{g} 
\end{equation}

Now we propose the  form of this function. Firstly, it must contain kinetic
term for tachyon, which should be written in manifest supersymmetry invariant
way:
\begin{equation}\label{kin}
I_{KT}=\tilde{\mathcal{G}}_S^{\mu\nu}\partial_{\mu}T\partial_{\nu}T
\end{equation}
where $\tilde{\mathcal{G}}$ denotes the matrix inverse of $\mathcal{G}+
\frac{1}{2\pi}\mathcal{F}$ and
$\tilde{\mathcal{G}}_S$ means the symmetric part of the matrix. As was argued
in ref.\cite{Sen}, the choice of this metric was motivated by work \cite{Witen2},
where was argued that in constant background $B$ field the natural metric for
open strings is $\tilde{\mathcal{G}}_S$ and ordinary products becomes noncommutative
products. On the other hand, in \cite{Chu1,Chu2}  it was shown, that natural noncommutative
parameter is gauge invariant combination of $F-B$, where $F$ is a constant background field strength of
gauge field living on world-volume Dp-brane. From these reasons, we take the
same metric as in \cite{Sen}, but due to the vanishing of $B$ field and background
$F$ field strength we expect that noncommutative effects do not appear in our
case and hence we will consider ordinary products between any functions.
We will see that the function $F$ is really invariant under all symmetries presented
above. 

There should be potential term for tachyon in function $F$. We suppose form
of this term as:
\begin{equation}\label{pot}
V(T)=-m^2T^2+\lambda T^4+ \frac{m^4}{4\lambda}
\end{equation}
This potential has a vacuum value equal to
\begin{equation}\label{vv}
\frac{dV}{dT}=0 \Rightarrow T_0^2=\frac{m^2}{2\lambda}
 , \  V(T_0)=0
\end{equation}

We also expect that some interaction terms between tachyon and $X$ fields
and gauge fields will be presented in $F$. In fact the interaction between
$T$ and $X, A$ is presented in kinetic term for tachyon, which can be seen
from the form of $\tilde{\mathcal{G}}_S$.  We must also stress that tachyon
is not charged with respect to gauge field because transforms in adjoin 
representation of gauge group and $U(1)$ has no adjoin representation. From
this reason there are no covariant derivatives in the action.

As a last thing, we will consider the interaction term between tachyon
and fermionic fields $\theta_L, \theta_R$. We propose this term in the form:
\begin{equation}
 I_{TF}=g(T)(\theta_R^T\Gamma^0\Gamma_M\Pi^M_{\mu}
\tilde{\mathcal{G}}_S^{\mu\nu}\partial_{\nu}\theta_R+
\theta_L^T\Gamma^0\Gamma_M\Pi^M_{\mu}
\tilde{\mathcal{G}}_S^{\mu\nu}\partial_{\nu}
\theta_L )+ f(T)\tilde{\mathcal{G}}_S^{\mu\nu}\partial_{\mu}\theta_R^T\Gamma^0
\partial_{\nu}\theta_L 
\end{equation}
where $g(T)$ is even function of $T$ and $f(T)$ is odd function of $T$, which
comes from the fact that in perturbative diagrams in string theory  $T$ comes
with CP factor $\sigma_1$ and $\theta_L$ with CP matrix ${\bf 1}$ and finally
$\theta_R$ with CP factor $\sigma_1$. Then it is clear that $\theta_R\theta_L $
gives $\sigma_1$ so in order to have nonzero trace over CP factors, we must
have $f(T)$ with odd powers of $T$, which gives factor $\sigma_1$. In the same
way it can be shown that $g(T)$ must be even function.
We have also used the important properties of
Majorana-Weyl spinors which say that expression of two Weyl spinors of the 
same chirality with odd number of gamma matrices is zero and expression of
two Weyl spinors of opposite chirality with even number of gamma matrices is zero,
which can be seen from following simple arguments:
\begin{equation}
\theta_L^T(\Gamma_i...\Gamma_k)\theta_L=\theta_L^T\Gamma_{11}
(\Gamma_i...\Gamma_k)\Gamma_{11}\theta_L=
-\theta_L^T(\Gamma_i...\Gamma_k)\theta_L
\end{equation} 
where we have used the fact, that $\Gamma_{11}$ commutes with even
number of gamma matrices and anticommutes with odd numbers of gamma 
matrices.

In summary, we expect that $F$ has a form:
\begin{equation}\label{F}
 F=\frac{2\pi\sqrt{2}}{g}(
\tilde{\mathcal{G}}_S^{\mu\nu}\partial_{\mu}T\partial_{\nu}T
+V(T)+I_{TF}) 
\end{equation}
We know that $F$ must be invariant under supersymmetric transformations
as well as under Lorenz transformations and translations. 
We will see that requirement of supersymmetric invariance place important conditions
on various terms in the action.

Under space-time translation, which has a 
form 
\begin{equation}
\delta_{\xi}X^M=\xi^M, \ \delta_{\xi} \theta_{L,R}=0, \ \delta_{\xi}T=0 
\end{equation}
we have
\begin{equation}
 \delta_{\xi}\Pi^M_{\mu}=0 \Rightarrow \delta_{\xi}\mathcal{G}=0 
, \ \delta_{\xi} \mathcal{F}=0
\end{equation}
so that $F$ is invariant.

Under $SO(1,9)$ Lorenz symmetry  various fields transform as
\begin{equation}
 X'^M=\Lambda^M_N X^N, \theta'=R(\Lambda)\theta, \overline{\theta}'=
\overline{\theta}R(\Lambda)^{-1}, T'=T 
\end{equation}
 We  than obtain
\begin{equation}
 \Pi'^M_{\mu}=\Lambda^M_N\partial_{\mu}X^N-\overline{\theta}R(\Lambda)
^{-1}\Gamma^M R(\Lambda)\theta=\Lambda^M_N\Pi^N_{\mu} 
\end{equation}
where we have used: $R(\Lambda)^{-1}\Gamma^MR(\Lambda)=\Lambda^M_N
\Gamma^N $.
Then 
\begin{equation}
\mathcal{G}'_{\mu\nu}=\eta_{MN}\Lambda^M_K\Pi^K_{\mu}
\Lambda^N_L\Pi^L_{\nu}=
\mathcal{G}_{\mu\nu}
\end{equation}
with using $\Lambda^M_K\eta_{MN}\Lambda^N_L=
\eta_{KL} $.
In the similar way we can show that $\delta_{\Lambda}
\mathcal{F}=0$ and consequently
\begin{equation}
\delta_{\Lambda}\tilde{\mathcal{G}}_S=0
\end{equation}
To prove Lorenz invariance of fermionic terms we use
\begin{equation}
\theta_R=\frac{1}{2}(1+\Gamma_{11})\theta, \theta_L=
\frac{1}{2}(1-\Gamma_{11})\theta 
\end{equation}
as well as basic properties of $\Gamma_{11}$ matrix:
$ \Gamma_{11}^T=\Gamma_{11} , R(\Lambda)^{-1}\Gamma_{11}
R(\Lambda)=\Gamma_{11} $
to rewrite the expression $\partial_{\mu}\theta_R\Gamma^0\partial_{\nu}
\theta_L $ as
\begin{equation}
\frac{1}{4}\partial_{\mu}\theta^T(1+\Gamma_{11})\Gamma^0
(1-\Gamma_{11})\partial_{\nu}\theta=\frac{1}{2}\partial_{\mu}
\overline{\theta}(1-\Gamma_{11})\partial_{\nu}\theta 
\end{equation}
which transforms under Lorenz transformation as
\begin{equation}
 \frac{1}{2}\partial_{\mu}\overline{\theta}'(1-\Gamma_{11})
\partial_{\nu}\theta'=\frac{1}{2}\partial_{\mu}\overline{\theta}
R(\Lambda)^{-1}(1-\Gamma_{11})R(\Lambda)\partial_{\nu}\theta=
\frac{1}{2}\partial_{\mu}\overline{\theta}\partial_{\nu}\theta 
\end{equation}
which prove the Lorenz invariance of this term.
In the same way we can prove invariance of expression 
\begin{equation}
 \theta_R^T\Gamma^0\Gamma_M\Pi^M_{\mu}\partial_{\nu}\theta_R=
\frac{1}{2}\overline{\theta}\Gamma_M(1+\Gamma_{11})\Pi_{\mu}^M
\partial_{\nu}\theta
\end{equation}
 This term
transforms under Lorenz transformation as
\begin{eqnarray}
 \frac{1}{2}\overline{\theta}R^{-1}\Gamma_M(1+\Gamma_{11})
\Lambda^M_K\Pi^K_{\mu}R\partial_{\nu}\theta=
\frac{1}{2}\overline{\theta}\Lambda_M^L\Gamma_L(1+\Gamma_{11})
\Lambda^M_K\Pi_{\mu}^K\partial_{\nu}\theta= \nonumber \\
=\frac{1}{2}\overline{\theta}\Gamma_M(1+\Gamma_{11})\Pi^M_{\mu}
\partial_{\nu}\theta
\end{eqnarray}
 We
see  that the all integration terms between fermions and tachyon are 
Lorenz invariant. 

Now we come to the crucial question of supersymmetry transformation,
which have a form
\[ \delta_{\epsilon}\theta=\epsilon, \delta_{\epsilon}X^M=
\overline{\epsilon}\Gamma^M\theta \]
It is well known, that these transformations leave $\Pi^M_{\mu}$ invariant,
consequently $\mathcal{G} $ as well. It can be also shown \cite{Aganagic}
that $\mathcal{F}$ is invariant as well. As a result we have
\begin{equation}
\delta_{\epsilon}\tilde{\mathcal{G}}_S=0
\end{equation}
Now we are ready to prove invariance of 
the term
\begin{equation}
 f(T)\tilde{\mathcal{G}}_S^{\mu\nu}\partial_{\mu}\theta_R^T\Gamma^0\partial_{\nu}
\theta_L=\frac{1}{2}f(T)\tilde{\mathcal{G}}_S^{\mu\nu}\partial_{\mu}
\overline{\theta}(1-\Gamma_{11})\partial_{\nu}\theta 
\end{equation}
This term is clearly invariant under supersymmetry transformations due to
the presence of partial derivative. On the other hand, the term
\begin{equation}
 \frac{1}{2}g(T)\tilde{\mathcal{G}}_S^{\mu\nu}
\overline{\theta}\Gamma_M(1+\Gamma_{11})\Pi^M_{\mu}\partial_{\nu}\theta 
\end{equation}
leads after supersymmetric transformation to the variation of the action
\begin{equation}
 \delta S=\int D \overline{\epsilon}\Gamma_M(1+\Gamma_{11})\Pi^M_{\mu}
\partial_{\nu}\theta= -\int \partial_{\nu}(D\Pi^M_{\mu})\overline{\epsilon}
\Gamma_M(1+\Gamma_{11})
\theta 
\end{equation}
where we have used
\begin{equation}
D=\sqrt{-\det (\mathcal{G}+\frac{1}{2\pi}\mathcal{F})}
\frac{1}{2}g(T)\tilde{\mathcal{G}}_S^{\mu\nu} 
\end{equation}
We see that requirement of invariance under 
transformation of supersymmetry leads to conclusion that
term $ g(T)... $ should not be present in the action for non-BPS D-brane, so that we
consider interaction term between fermions and tachyon in the form:
\begin{equation}
 f(T)\tilde{\mathcal{G}}_S^{\mu\nu}\partial_{\mu}
\theta_R^T\Gamma^0\partial_{\nu}\theta_L
\end{equation}
It is important to stress that this term is consistent with requirement, that $f(T)$ should
be odd function of $T$. This can be seen from the fact that $\tilde{\mathcal{G}}_S$
has  a CP factor equal to unit matrix. To prove this we expand
 $\mathcal{G}$ as follows:
\begin{equation}
\mathcal{G}_{\mu\nu}=\eta_{MN}\partial_{\mu}X^M\partial_{\nu}X^N
-2\eta_{MN}\partial_{\mu}X^M\overline{\theta}\Gamma^N\partial_{\nu}\theta+
\eta_{MN}(\overline{\theta}\Gamma^M\partial_{\mu}\theta)
(\overline{\theta}\Gamma^N\partial_{\nu}\theta)
\end{equation}
we know that $X^M$ comes from CP sector with unit matrix, so that only one
"dangerous" term is
\begin{equation}
\overline{\theta}\Gamma^N\partial_{\nu}\theta=
(\theta_R+\theta_L)^T\Gamma^0\Gamma^N\partial_{\nu}
(\theta_R+\theta_L)=
\theta_R^T\Gamma^0\Gamma^N\partial_{\nu}\theta_R+
\theta_L^T\Gamma^0\Gamma^N\partial_{\nu}\theta_L 
\end{equation}
These two terms give CP factors either $(\bf 1\bf 1={\bf 1})$ (for $\theta_L$)
or $(\sigma_1\sigma_1)={\bf 1}$ for ($\theta_R$)  (In previous part 
we have used the result:
 $\theta_R^T\Gamma^0\Gamma^M\partial_{\mu}\theta_L=-
\theta_R^T\Gamma_{11}\Gamma_0\Gamma^M\Gamma_{11}\partial
_{\mu}\theta_L=-\theta_R^T\Gamma^0\Gamma^M\partial_{\mu}\theta_L$).

In the same way we can prove that $\mathcal{F}$ comes with unit matrix
of CP factors. For example expression $ \overline{\theta}\Gamma_{11}
\Gamma_M\partial_{\mu}\theta $ is equal to
$ \theta_R^T\Gamma^0\Gamma_{11}\Gamma_{M}\partial_{\mu}\theta_R
+\theta_L^T\Gamma^0\Gamma_{11}\Gamma_{M}\partial_{\mu}\theta_L$,
where we have used the identity
\begin{equation}
\theta_L^T\Gamma^0\Gamma_{11}\Gamma_M\theta_R=
-\theta_L^T\Gamma_{11}\Gamma_0\Gamma_{11}
\Gamma_M\Gamma_{11}\theta_R=
-\theta_L^T\Gamma_0\Gamma_{11}\Gamma_M\theta_R
\end{equation}

It seams to us that requirement of supersymmetry places strong constraint on
various coupling between fermions and tachyon. In particular, 
we have seen that fermions must always come with partial derivative. In next
section we will shown that tachyon condensation in form of kink solution
leads to correct supersymmetric invariant DBI action of BPS D(p-1)-brane.

\section{Tachyon condensation on world-volume of non-BPS
D-brane}
In this section we will consider tachyon condensation on
world-volume of non-BPS Dp-brane in the form of kink solution.
In order to get clear picture of resulting D-brane, we will consider
tachyon kink solution in the form:
\begin{equation}\label{sol}
T(x)=\left\{ \begin{array}{ccc} -T_0 , \ x< 0 \\
                                                0 , \ T=0 \\
                                             T_0 , \ x>0 \\
\end{array}\right.
\end{equation}
where $x$ is one particular coordinate on world-volume of non-BPS
Dp-brane. We get equation of motion for tachyon from variation
of (\ref{action}), which give (we consider dependence of tachyon only
on $x$, say $p$ coordinate):
\begin{equation}\label{eq}
\frac{d}{dx}\left(\frac{\delta F}{\delta \partial_x T}\right)
-\frac{d F}{dT}=0
\end{equation}
where $F$ has a form:
\begin{equation}\label{F2}
F=\frac{2\pi\sqrt{2}}{g}\left(
\tilde{\mathcal{G}}_S^{\mu\nu}\partial_{\mu}T\partial_{\nu}T
+V(T)+f(T)\tilde{\mathcal{G}}^{\mu\nu}_S
\partial_{\nu}\theta_R^T\Gamma^0\partial_{\nu}\theta_L\right) 
\end{equation}

First equation in (\ref{eq}) gives
\begin{equation}
\partial_{\mu}(\tilde{\mathcal{G}}^{\mu x}_S\partial_x T(x) )
=\partial_{\mu}(\tilde{\mathcal{G}}^{\mu x})\partial_x T=0
\end{equation}
where we have used the fact that tachyon field is only
function of $x$ and the fact that we can look at the solution
(\ref{sol}) as a extreme limit of ordinary kink solution, when
behaviour of tachyon field can be approximated around the
point $x=0$ as $T \sim x$. More precisely, when we take solution
which is equal to tachyon vacuum value $T_0$ at distance $x=L$
from the point $x=0$ than we get tachyon field around the point
$x=0$ in the form $T=\frac{T_0}{L}x$. It is clear that second
derivative of this expression is zero.  In order to get (\ref{sol}) we
must take limit $L\rightarrow 0$ for $x\rightarrow 0$ in such a way
that $T\rightarrow 0$ for $x\rightarrow 0$. 

Outside the point $x=0$ this equation is trivially obeyed because
$T(x)$ is equal to its vacuum value $T_0$. In the point $ x=0$ we
have $\partial_x T(x) \sim \delta(x) $ so that to obey equation of
motion we must pose the condition
\begin{equation}\label{con}
\left.\tilde{\mathcal{G}}^{\mu x}_S\right |_{x=0}=const.
\Rightarrow \mathcal{G}_{\mu x}\left |_{x=0}\right.
=const. , \ \mathcal{F}_{\mu x}\left|_{x=0}\right.=
const.
\end{equation}
When we look at definition of $\mathcal{G}$ we will see, that
this condition implies $X^{\mu}=k x$ which has a form of 
static gauge which certainly is not correct for $\mu \neq x$
so we  come to the result $\mathcal{G}_{\mu x}=0$
for $\mu \neq x$ and with appropriate form of scaling we
can take $\mathcal{G}^{xx}=\mathcal{G}_{xx}=1 $.
 For $\mathcal{F}_{\mu x}
=F_{\mu x}=const. $ (where we have used the upper 
result $\partial_x X^{\mu}=\partial_x \theta_L=
\partial_x \theta_R=0, \ \mu \neq x$) we can take this constant
to be equal to zero, because the presence of the constant term
do not affect the equation of motion.

The second term in (\ref{eq}) gives 
\begin{equation}
-\frac{dV}{dT}-\frac{df(T)}{dT}\tilde{\mathcal{G}}_S^{\mu\nu}
\partial_{\mu}\theta_R^T\Gamma^0\partial_{\nu}\theta_L=0
\end{equation}

It easy to see that (\ref{sol}) is a solution
of equation $\frac{dV}{dT}=2m^2T(
\frac{T^2}{T^2_0}-1)=0$. Outside the point $x=0$ the tachyon field
is equal to its vacuum value so the expression in the bracket is zero and
in the point $x=0$,  $T$ is equal to  zero. In the point $x=0$ the
value of potential is equal to $\frac{m^4}{4\lambda}$. 

Now we come to the second term in (\ref{eq}). We know that
$f(T)$ is odd function of $T$ so that we can expect that  its derivative $\frac{df(T)}{dT}$
is nonzero. The only possibility to obey equation of motion for tachyon is
to pose the condition that $\partial_{\mu}\theta_R $ or 
$\partial_{\mu}\theta_L$ is equal to zero. We choose the
condition:
\begin{equation}\label{fermi}
\partial_{\mu}\theta_R=0
\end{equation}
Strictly speaking we have obtained this condition from
behaviour of tachyon outside the point $x=0$. In the point $x=0$ 
we expect that derivative of $f$ is still nonzero but 
$\tilde{\mathcal{G}}_S^{\mu x}=0$ so that in the point $x=0$
we have condition
\begin{equation}
\partial_{\mu}\theta_R=0 , \ \mu=0,...,p-1
\end{equation}
 but with using (\ref{con}) and consequences resulting from
it we see that (\ref{fermi}) must holds in the point $x=0$ as well.
We have arrived to the important result that $\theta_R$ is 
non-dynamical field which can be set to equal to zero. In other words,
we have eliminated via tachyon condensation one half of fermionic
degrees of freedom with direct parallel of gauging of $\kappa $ symmetry on
world-volume of BPS D-brane. 

We come to the final result. We have seen that (\ref{sol}) is equation
of motion for tachyon. When we put this function into (\ref{F2}) we 
get
\begin{equation}\label{Fresult}
F(T=kink)=\frac{2\pi\sqrt{2}}{g}T^2_0\delta (x)
\end{equation}
where we have used $\tilde{\mathcal{G}}_S^{xx}=1$ and $\delta(x)^2=\delta(x)$.
In previous equation we have also used the fact that the derivation is much
larger that the vacuum value of  the potential. In order to get correct tension of resulting D-brane we must fix
the vacuum value of tachyon to be equal to $T^2_0=\frac{1}{\sqrt{2}}$.
When we put (\ref{Fresult}) into (\ref{action}) and with using (\ref{con}), 
(\ref{fermi})
we get the $\kappa$ fixed action for D(p-1)-brane in 
Type IIA theory:
\begin{equation}
S=-\frac{2\pi}{g}\int d^{p}\sigma 
\sqrt{-\det(\mathcal{G}+\frac{1}{2\pi}\mathcal{F})}
\end{equation}

There is a place when we can discuss some questions
about interaction terms between fermions and tachyon. In 
previous part we have estimated the possible form of 
interaction term between fermions and tachyon on grounds
of invariance under supersymmetry. We may ask the question,
whether other possible terms which were not included into the
form of $F$ do not spoil the result of emergence of BPS D-brane
from tachyon condensation on world-volume of non-BPS D-brane.
For example, we can consider the interaction of the form:
\begin{equation}
\tilde{\mathcal{G}}_S^{\rho\kappa} \partial_{\rho}g(T^2)\Pi^M_{\kappa}
\tilde{\mathcal{G}}_S^{\mu\nu}\partial_{\mu}\overline{\theta}_L\Gamma_M
\partial_{\nu}\theta_L 
\end{equation}

which is consistent with requirement of invariance under all
symmetries presented above as well as with the trace of  CP factors.
 The answer is  that this term is zero for tachyon approaching
kink solution and consequently do not give any constraint on 
massless fields.

When we write $g(T)=T^2+T^4...$, than
$\partial_x g(T)=2T\partial_x T+4T^3\partial_x T+...$ and consequently
$\frac{\delta \partial_x g}{\delta\partial T}=2T+4T^3+...$ so that the first term
in equation of motion for tachyon gives contribution
\begin{equation}
 2\partial T+ 12T^2\partial T+...
\end{equation}
while the second term is equal to
\begin{equation}
 2\partial T+12T^2\partial T+...
\end{equation}
then we see, that the term  $\partial_x g(T^2)$ obeys equation of motion identically and we do
not get any additional constraint on massless fields. Moreover, when
we put kink solution into this term we will see that this term is identically equal to
zero. Outside the core of the soliton $T$ is
in its vacuum value $T^2=T^2_v$ so that its derivative is zero and in the
point $x=0$ we can consider this kink solution as a extreme limit
of ordinary solution. Than $T\partial T$ is zero in the point $x=0$ due to the fact
that 
$T$ is zero in the point $x=0$. When we go to extreme limit, this expression
is always equal to zero  in this limiting procedure so we can conclude
that is zero in extreme limit as well. We than see that this interaction
term is identically zero after tachyon condensation and do not give
 any new information and any new constraint on fermionic field so
that we do not include this term into definition of $F$ function in (\ref{action}).
We also expect that other possible interaction terms do not 
give new information and from this reason we will consider the $F$ function
in the form
 given in (\ref{F2}).

We must also discuss the situation when $T$ is equal to its vacuum value
everywhere. Naively we could expect from the form of $(\ref{F2})$ that
for this value of tachyon we do not get supersymmetric vacuum due to
the presence of the interaction term between tachyon and fermions. However,
tachyon vacuum value must be solution of equation of motion and as 
we have seen this leads to requirement of constant spinor field $\theta_R$.
Then interaction term between fermions and tachyon is equal to zero and
from the definition $V(T=T_0)=0 , \ \partial_x T=0$ and we get the result that
the second bracket is equal to zero and consequently the whole action
disappears with agreement with ref.\cite{Sen}.

\section{Conclusion}

In previous parts we have proposed the possible form of supersymmetric 
DBI action for non-BPS D-brane in Type IIA theory (For IIB theory the
situation will be basically the same with difference that both spinors have the
same chirality). We have seen that requirement of invariance under
supersymmetric transformations place strong constraints on possible form
of this action.  Then we have studied the kink solution of tachyon on 
world-volume of non-BPS Dp-brane in IIA theory and
we have shown, that this solution really describes BPS D(p-1)-brane in IIA theory, which is in agreement with 
results in
\cite{Sen1,witen,Horava} and in some sense can serve  as further support of their results. 

We would like to outline the possible extension of this work. It would be certainly nice to study situation when we have
$N$ non-BPS D-branes and tachyon condenses in more general configuration.
It would be also interesting to study tachyon condensation on system of D9-branes
and antibranes in Type IIB theory following \cite{witen}. We hope to return to
these important question in the future.

\end{document}